\documentclass{svproc}
\usepackage{url}
\usepackage{graphicx}
\usepackage{footmisc}
\usepackage{amsmath}

\begin{document}
\mainmatter              
\title{Agentization of Two Population-Driven Models of Mathematical Biology}
\titlerunning{Agentization of Population-Driven Models}  
%
\author{John C. Stevenson \\ \tiny jcs@alumni.caltech.edu}
%
%
%
\institute{}

\maketitle              

\begin{abstract}
	
Single species population models and discrete stochastic gene frequency models are two standards of mathematical biology important for the evolution of populations. An agent based model is presented which reproduces these models and then explores where these models agree and disagree under relaxed specifications. For the population models, the requirement of homogeneous mixing prevents prediction of extinctions due to local resource depletion. These models also suggest equilibrium based on attainment of constant population levels though underlying population characteristics may be nowhere close to equilibrium. The discrete stochastic gene frequency models assume well mixed populations at constant levels.  The models' predictions for non-constant populations in strongly oscillating and chaotic regimes are surprisingly good, only diverging from the ABM at the most chaotic levels.

\end{abstract}
\section{Introduction}
Mathematical biology provides many models of single species population dynamics that are important for evolutionary processes, two of which are: single species logistic equations (both continuous and discrete), and  discrete stochastic models of gene frequencies in finite populations. The discrete logistic equations generate oscillating and chaotic population dynamics not seen in the continuous model, and allow for some extinctions \cite{murray}. The discrete stochastic gene frequency models are important for animal husbandry, trait extinction, and, with some of these models, retrospective studies of allele frequency distributions based on current conditions. \cite{ewens}.

These models represent the "struggle for existence" which is the driver of evolution \cite{gause}. Any agent based model that evolves agent or population characteristics should reproduce these standard results.  An ABM is presented that reproduces these models' results and then compares the behaviors of these standard models to the ABM when the specifications are relaxed. The ABM modeling of the logistic growth equations reveals two important behaviors not seen in the standard models: spatially driven local extinctions and non-equilibrium population characteristics extending orders of magnitude past the time that the population level itself achieved equilibrium. The ABM modeling of gene frequencies reproduces the predictions of fixation probability, and mean absorbtion times of the standard discrete stochastic models across a wide range of intrinsic growth parameters, up to the edge of chaotic population trajectories. These parameters are investigated for both neutral and weak selection pressures.

The ABM is based on the model of Epstein and Axtell \cite{eps:axl}. The model uses identical agents on an equal opportunity (flat) landscape to match the standard models. For the logistic growth equations, as in the study of bacteria \cite{gause}, the population trajectory begins with a single agent and is dependent on a stochastic growth parameter: infertility. There is a one-to-one mapping of the infertility parameter of the ABM with the intrinsic growth rate of the single species logistic growth equations. Detailed descriptions of the ABM parameters and process are given in Appendix A. The mathematics of continuous and discrete logistic growth equations used in this study are given in Appendix B.

This ABM, with the addition of a single gene with two alleles, allows replication of the Wright Fisher class of discrete stochastic models of gene frequency. The probability of allele fixation and mean absorption times are replicated for the Cannings \cite{cannings} and Moran \cite{moran} models of two allele haploid populations of fixed size. These models are currently important in animal husbandry, ecological studies of extinction, and anthropological studies of past allele diffusion based  on the current populations \cite{ewens}.

\section{Discrete Logistic Equation with Time Delay}

The fields of mathematical biology and ecology developed equation-based continuum modeling of single species populations, models both continuous and discrete \cite{murray} \cite{kot}. A continuous homogeneous model of single species population was proposed by Verhulst in 1838 \cite{verh} and allows an exact solution for the intrinsic growth rate. While the continuous model fits the initial phase of growth well, it does not model oscillating population levels at high rates of intrinsic growth. These types of periodic and chaotic oscillations are often generated by population models that are discrete, that contain time delays, or both \cite{liz}. Researchers in the fields of biology and ecology have used these discrete and delayed population models to handle, for example, species that have no overlap between generations \cite{murray} or have specific breeding seasons \cite{kot}. To account for such delays in animal populations, Hutchinson in 1948 \cite{hutch}  and Wright in 1955 \cite{wright} extended the Verhulst process, now often referred to as the Hutchinson-Wright equation, as \cite{kot} 

\begin{equation}
	N(t+1)=[1+r-\frac{N(t-\tau)}{K}]N(t)
\end{equation}
with an explicit time delay $\tau$ in the self-limiting term. For the ABM, the time delay captures both the landscape cell's regeneration of resources and the limited range of the agents' movements and vision.

Figure 1 shows the population trajectories generated by the ABM with specified infertilities; and the continuous Verhuslt and discrete Hutchinson-Wright (1) trajectories with appropriate intrinsic growth rates and time delay. The regimes of these trajectories move from stable on the right, to steady oscillations in the middle, to fully chaotic on the left based on increasing growth rates.

\begin{figure}
	\begin{center}
		\includegraphics[angle=-90,scale=0.65]{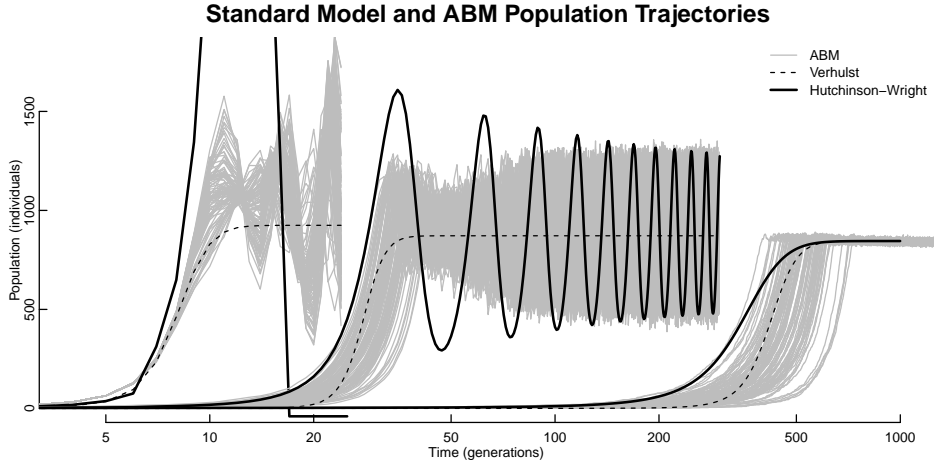}
	\end{center}
	\caption{Population Trajectories} {ABM population trajectories for infertility 1, 5 and 85 (from left to right) with best fit Verhulst intrinsic growth rate (10) and Hutchinson-Wright (1) intrinsic growth rate and delay coefficient}
\end{figure}

\subsection{Relaxation of Specifications - Extinction}

These single species population models assume a well mixed, spatially homogeneous populations. Furthermore, only the number of individuals in the population over time is modeled. The replicating ABM also provides both spatial resolution and addition descriptors of the populations. These measures provide significant additional insights into the limits of the standard models and will be addressed in turn.

While the discrete logistic equation does generate extinctions, these extinctions occur for $ N_{t+1}<0$ and are due to exceptionally large values of $N_{t}$ as shown by the Hutchinson-Wright trajectory at the far left of Figure 1. The logistic equation is often modified (Ricker Curve) to avoid these "unrealistic" excursions \cite{murray}. The ABM with its spatial modeling as opposed to the well mixed specification of the standard model does not immediately go extinct for these cases.

In regimes of large oscillations, the replicating ABM generates different extinctions based on a local lack of resources even though the global resources are more than adequate. These extinctions are driven by spatial waves of agents who consume everything in their path until they find themselves in a local desert of resources, unable to see or move to the ample food. Whether the population becomes extinct is dependent on a stochastic process where a few individuals may survive behind the wave to seed the next generation. With such a dependency on a stochastic process, extinction is inevitable. Figure 2 shows both events for the ABM: the top row where a few individuals survive behind the wave and the bottom row where none survive and extinction occurs in a local desert.

\begin{figure}
	\begin{center}
		\includegraphics[angle=-90,scale=0.65]{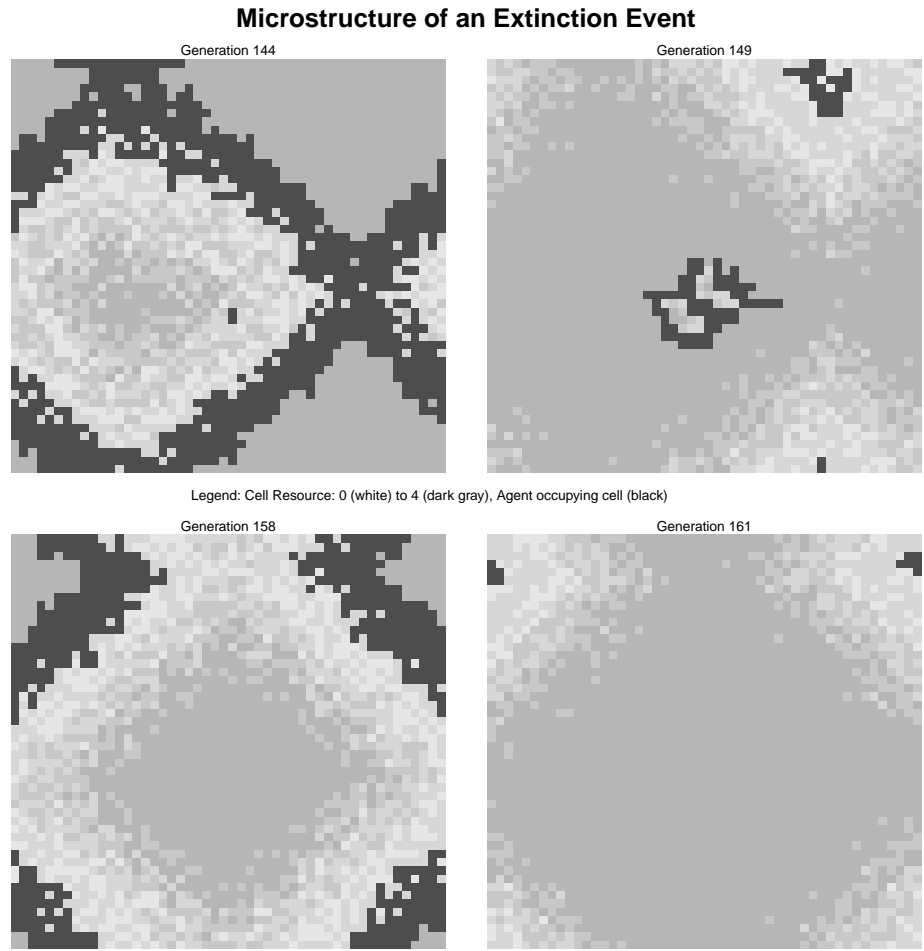}
	\end{center}
	\caption{Spatial Extinction} {a) When the penultimate wave of agents pass through the rich landscape, a few agents manage to survive behind the wave (Generation 144). As the last agents of the wave perish in a local desert, the next generation blooms (Generation 149). For the final wave, no agents survive behind the wave as it passes (Generation 158). The last cycle finds the remaining agents all in a local desert and about to perish (Generation 161). This extinction event is driven by non-equilibrium movement dynamics, and spatially local resource densities. Together these factors produce a highly stochastic process leading to certain extinction but at an unpredictable time.}
\end{figure}

\subsection{Relaxation of Specifications - Equilibrium}

Equilibrium in these population models is considered to occur when the population level has reached a steady level in the stable growth regime of the standard model. The ABM has additional measures of the population available when replicating these cases that reveal an underlying lack of equilibrium as measured by population age. Figure 3 demonstrates this mismatch in equilibrium times for both low growth and high growth (but not oscillatory) configurations. 
\begin{figure}
	\begin{center}
		\includegraphics[angle=-90,scale=0.65]{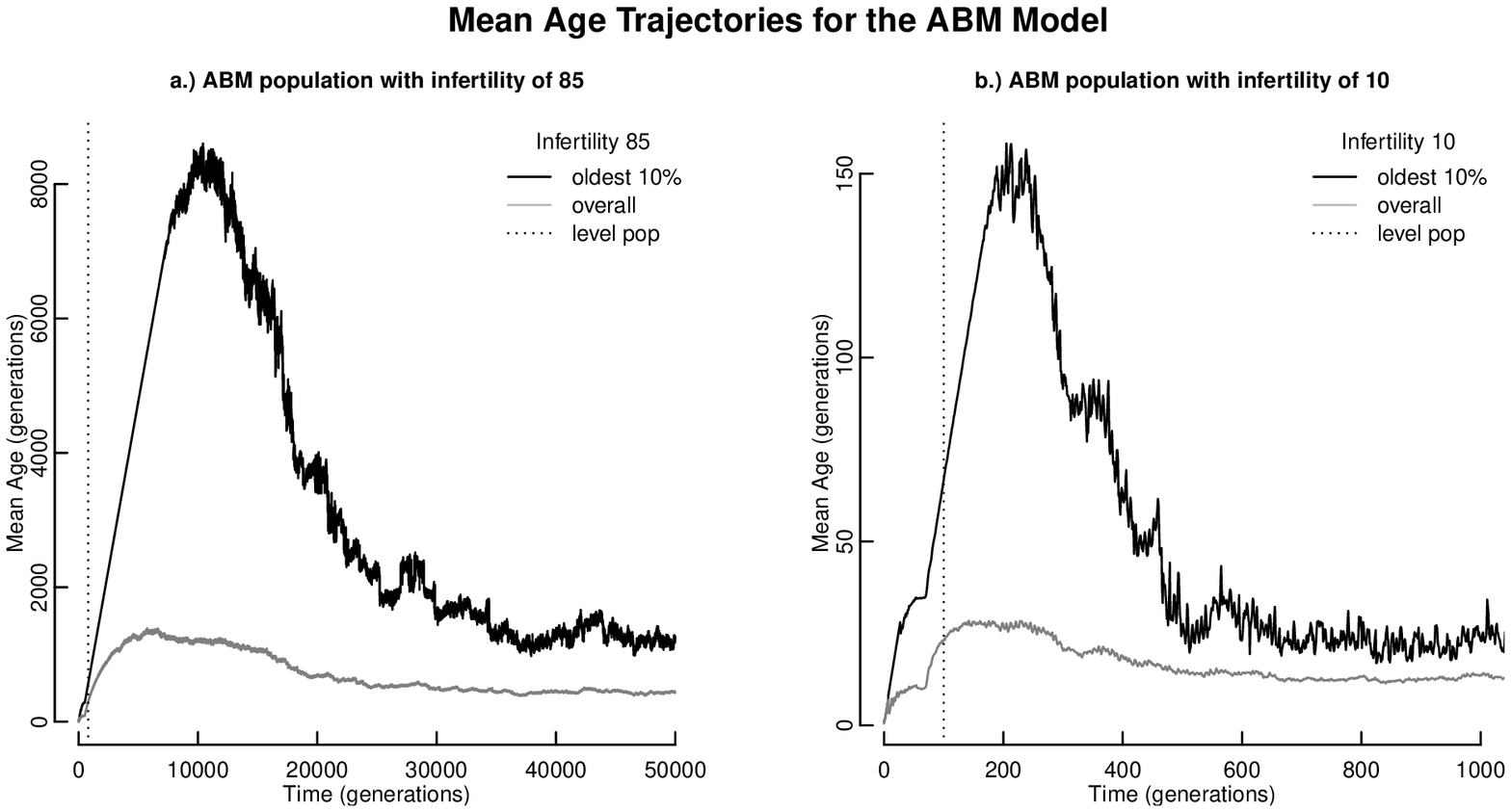}
	\end{center}
	\caption{Mean Age Equilibrium} {ABM mean age trajectories for infertility of 85 (left) and 10 (right). Vertical dotted lines show the generation at which the population level reach equilibrium.}  
\end{figure}

These additional characteristics do not achieve equilibrium, sometimes for orders of magnitude greater than for the population level to reach equilibrium. The impact of this dynamic is significant for results based on these underlying metrics, for example wealth inequality \cite{stevenson}
 
\section{Models of the Stochastic Behavior of Genes in Finite Populations} 
 
An important generalization of the Wright-Fisher class of models was introduced by Cannings \cite{cannings}. This model allows overlapping generations as seen in the ABM models of populations. Its stochastic rule is based on the distribution of descendant genes. The second model considered was introduced by Moran \cite{moran} and has the advantage of providing "explicit expressions for many quantities of evolutionary interest" and is reversible, allowing for retrospective studies \cite{ewens}.
 
The specific configuration of these standard models chosen for this study is a haploid population with one gene and two alleles, no mutations, and, initially no selective pressure. The standard models assume a constant population, a specification which will be relaxed. For the Cannings model, the population at each $t+1$ generation is based on the random possibility of reproduction, and survival or death of each individual, over the entire population. For the Moran model, at generation $t+1$, one individual is selected for reproduction and one individual is then randomly selected to die, excluding only the newborn. 

The predictions used for this study are the probability of allele fixation, the mean time to absorption of a single allele in an otherwise homogeneous population of the other allele (of interest for survival of a single mutation), and the mean time to absorption for initial equal number of alleles. The specification of constant population and homogeneous mixing are then relaxed and the results discussed.

For all the Wright-Fisher class of models with no mutations or selective pressure, the probability $\pi$ of one of the two alleles $(A_{1},A_{2})$ overtaking the whole population (fixation) is given as:
\begin{equation}
	\pi=i/2N
\end{equation}
where $i$ is the initial population of one of the two alleles and the total population is $2N$. The selection of $2N$ for the haploid population allows comparison with diploid populations of standard Wright-Fischer models. Figure 4a shows this simple linear relationship as a function of $i$ and the replication of these probabilities across the starting allele fraction for the ABM.
 
\begin{figure}
	\begin{center}
		\includegraphics[angle=-90,scale=0.65]{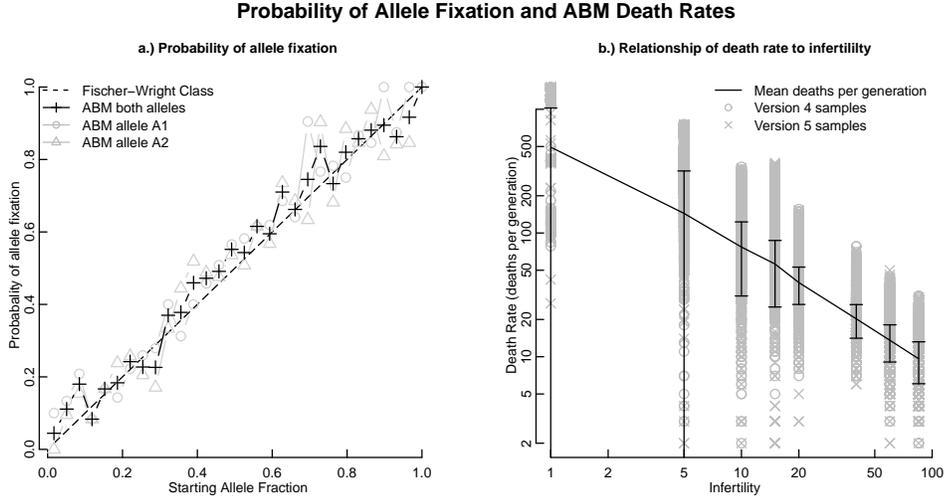}
	\end{center}
	\caption{Allele fixation and Death Rates} {a.) Comparison of ABM generated allele fixation probabilities versus allele starting frequency with the Wright-Fischer class of stochastic gene diffusion models. b.) Relationship of ABM infertility to death rates to support the Cannings and Moran models.}  
\end{figure}

Figure 4b demonstrates the one-to-one relationship of the ABM infertility parameter  with the ABM-generated death rates. This relationship is used to develop the Cannings and Moran models. The Cannings rule for stochastic reproduction of an individual $a_{i}$ at time $t$ can be expressed as vector $y(t)_{i}$ 
\begin{equation}
	y(t)_{i} = [2,2,1,0,0,1, ... 0 ]
\end{equation}
where $y_{i}$ ranges across all the individuals $a_{i}$ from $i=1$ to $i=2N$. The individual entries are the net population result for the descendants of each individual for that generation. Only one birth per individual per cycle is allowed.
The variance, $\sigma(t)^{2}$, of $y(t)_{i}$ is
\begin{equation}
	\sigma(t)^{2} =  \frac{1}{2N-1}\sum_{i=1}^{2N}[y(t)_{i}-\bar y(t)]^{2}
\end{equation}
where $\bar y(t)$ is mean of $y(t)_{i}$. With a constant sized population $\bar y(t)=1$, the births equal the deaths $d(t)$ , and (4) becomes
\begin{equation}
	\sigma(t)^{2} =  \frac{1}{2N-1}[\sum_{1}^{d(t)}(2-1)^{2}+\sum_{1}^{2N-2d(t)}(1-1)^{2}+\sum_{1}^{d(t)}(0-1)^2]
\end{equation}
where the number of individuals that survived but did not reproduce is equal to the total population minus those that gave birth minus those that died ($2N-2d$). Simplifying
\begin{equation}
	\sigma(t)^{2} =  \frac{1}{2N-1}[d(t)+0+d(t)]= \frac{2d(t)}{2N-1}
\end{equation}
The ability to calculate this variance directly for the ABM is quite useful. With this variance, the mean time to allele absorption $\bar t_{abs}$ is given as \cite{ewens}

\begin{equation}
	\bar t_{abs}= -(4N-2)[p\log(p)+(1-p)\log(1-p)]/\sigma(t)^{2}
\end{equation}
where p is the ratio of the number of one allele $A_{1}$ to the total population.

This general Cannings result can also produce the mean absorption times for Wright-Fisher with $\sigma^{2} \approx 1$ and Moran $\sigma^{2} \approx 2/(2N) $. Note for Moran the generation $t_{m}$ is a single individual's reproduction event and conversion to the Cannings $t_{c}$ is written as  
\begin{equation}
	t_{c} = t_{m}/2N
\end{equation}

The mean time to absorption for a single mutation $A_{1}$ in a population of $A_{2}$ is the mean time until the allele is lost, most likely, or takes over the population (unlikely but does occur with exceptionally long absorption times). Figure 5a gives the mean absorption times for a single allele, using the Cannings and Moran models from equation X and the results from the ABM runs. Figure 5b gives the results for these two standard models and the ABM over an even distribution of $p$.  
\begin{figure}
	\begin{center}
		\includegraphics[angle=-90,scale=0.65]{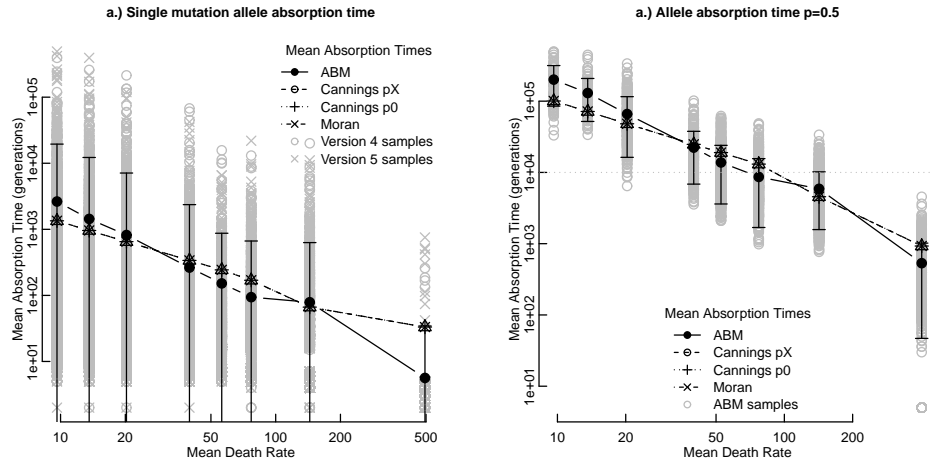}
	\end{center}
	\caption{Mean Allele Absorption Times} {a.) The ABM, Cannings, and Moran mean absorption times for a single mutation allele. b.) The allele absorption times for ABM , Cannings, and Moran models for an allele ration of 1/2.}  
\end{figure}

\subsection{Relaxation of Specification - Constant Population}
As can be seen in Figure 1, the population at steady state can only be considered constant for the dynamic regimes that are not oscillating nor chaotic. Surprisingly, as can be seen in Figures 5a and 5b, the standard models are in general agreement with the ABM for all but the most extreme chaotic regimes.

\subsection{Weak Selection}
Weak selection pressures are introduced to the ABM by modifying the agents' rule for movement towards the largest and the closest resources within range. If there are two different cells within range with the same largest resources and the same distance, the neutral ABM randomly selects between the ties. For the weak selection ABM, these ties are resolved based on one of the two alleles the agent carries: an introvert allele selects the target cell that has fewer other agents around it, and an extrovert allele that selects the target cell that has more agents surrounding it.
 
Using the Moran model\cite{ewens}, the probability the next individual selected to die is $A_{1}$ is 
\begin{equation}
\mu_{1}i/\{\mu_{1}+\mu_{2}(2N-i)\}
\end{equation}
where $i$ is the number of $A_{1}$ alleles in the $2N$ population and $\frac{\mu_{1}}{\mu_{2}}$ defines the selective advantage. If $\mu_{1}=\mu_{2}$ there is no selective advantage whereas if $\mu_{1}<\mu_{2}$ then allele $A_{1}$ has a small selective advantage. The probability of fixation $\pi$ is now:
\begin{equation}
	\pi_{i} = \{1-(\mu_{1}/\mu_{2})^{i}\}/\{1-(\mu_{1}/\mu_{2})^{2N}\}
\end{equation}
By defining $\mu_{1}/\mu_{2} = 1 - s/2$ with $s$ small and positive, Equation (10) can be approximated as
\begin{equation}
	\pi(x) = \{1-\exp(-\alpha x/2)\}/ \{1-\exp(-\alpha/2)\}
\end{equation}
where $x=i/2N$ and $\alpha = 2Ns$. With this approximation, the magnitude of the selection pressure as represented by $s$ can be implied from the ABM as on the order of 0.08, a weak selection pressure. Figure 6 a) presents the probability of fixation of the introvert and neutral phenotypes as a function of its initial ratio and the Moran standard model predictions of two of the implied parameters for weak and neutral selection pressure. Figure 6.b) presents the dramatically reduced absorption times. The single mutation fixation and absorption times diverge from the standard models predictions due to the stochastic effects on a single gene overriding the weak selection pressure\cite{ewens}.
\begin{figure}
	\begin{center}
		\includegraphics[angle=-90,scale=0.65]{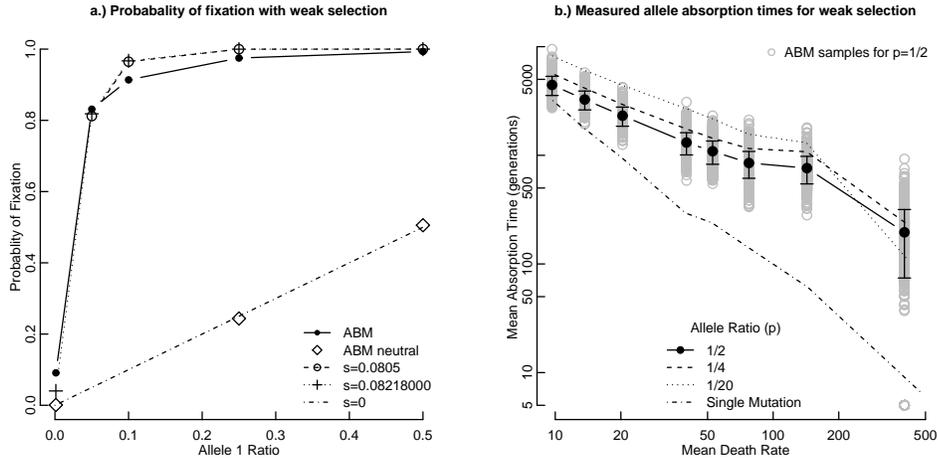}
	\end{center}	
	\caption {Weak Selection Fixation and Absorption} {a.)The probability of fixation of $A_{1}$. b) The mean time to absorption versus mean death rate for an allele ratio of 1/2.}
\end{figure}
\begin{center}
\end{center}

\section{Conclusions}

The ABM reproduced the results of the standard models well in the regimes that matched their specifications and showed interesting behavior when the specifications were relaxed.
For the Hutchinson-Wright time-delayed logistic model of population growth, the assumption of a well mixed population without a spatial dimension fails to predict extinctions due to local (spatial) resource depletion. This model also gives no indication of when the population reaches a steady state since the underlying characteristics such as mean age, may take many orders of magnitude more time to reach equilibrium.
The discrete stochastic models of gene frequency, on the other hand, survived well into the oscillating population level regimes and only differed from the ABM at the highest levels of chaotic regimes.

%
%

\pagebreak
\appendix{ \textbf{ Appendix A - Computational Model and Process}}

Table 1 provides the definition of the agents' and landscape's parameters used for this study. Vision and movement are along rows and columns only. The two dimensional landscape wraps around the edges (often likened to a torus). Agents are selected for action in random order each cycle. The selected agent moves to the closest visible cell with the most resources with ties resolved randomly. After movement, the agent harvests and consumes (metabolizes) the required resources. At this point, if the agent's resources are depleted, the agent is removed from the landscape. Otherwise an agent of sufficient age (not a newborn) then considers reproduction, requiring a lucky roll of the fertility die (infertility), and an empty von Neumann neighbor cell, which are only the four neighboring spaces one step away by row or column. The newborn is placed in the empty cell and remains inactive until the next action cycle. With this approach for the action cycle, no endowments are required whether for new births or for the agent(s) at start-up. Once all the agents have cycled through, the landscape replenishes at the growth rate and the cycle ends.

\begin{table}[h!]
	\begin{center}
	\begin{tabular}{|c|c|c|c|c|}
	\hline
			Agent Characteristic & Notation & Value & Units & Purpose \\
			\hline
			vision & $v$ &  6 &  cells & vision of resources on landscape \\
			movement & -- &  6 &  cells per cycle &  movement about landscape \\
			metabolism & $m$ & 3 & resources per cycle &  consumption of resource \\
			birth cost & $bc$ & 0 & resources &  sunk cost for reproduction \\
			infertility & $f$ & 1-85 & 1/probability & likelihood of birth \\
			puberty & $p$ & 1 &  cycles &  age to start reproduction\\
			surplus & $S$ & 0+ & resources &  storage of resource across cycles \\
			\hline
	\end{tabular}
	\bigbreak
	\begin{tabular}{|c|c|c|c|}
		\hline
		Landscape Characteristic & Notation & Value & Units\\
		\hline
		rows & -- & 50 & cells \\
		columns & -- & 50 & cells\\
		max capacity &$R$ & 4 & resource per cell\\
		growth & $g$ & 1 & resource per cycle per cell \\
		initial & $R_{0}$ & 4 & resource, all cells\\
		\hline
	\end{tabular}
	\caption{Agent and Landscape Parameters of the ABM}
	\label{Table 1:}
	\end{center}
\end{table}

The metabolism rate, uniform across a given population, is one that consumes per cycle 25\% less than the maximum capacity per cell.

\textbf{Control Volume Analysis}


A control volume is defined to track the flow of resources from generation on the landscape to destruction by the agents through metabolism, death, and birth. Note with no endowments, the only source of resources is growth on the landscape. Equations external to the ABM model are written below that provide verification, quantitative descriptions for replication, and analytic tools which reveal the internal workings of the ABM.

The change in resources within the landscape $\Delta E_{L}(t)$ per cycle can be written as
\begin{equation}
	\Delta E_{L}(t)= \sum_{c=1}^{N_{c}}g_{c}(t-1)-H(t)
\end{equation}
at the start of generation $t$, where $N_{c}$ are the landscape cells, and the growth rate for each of the cells in the landscape $g_{c}(t)$ is given as:
\begin{equation}
g_{c}(t) =  
  \begin{cases}
	g & r[c,t]+g\le R\\
	R-r[c,t] & r[c,t]+g>R\\
  \end{cases}
\end{equation}
with $R$ the maximum resources per cell, $g$ the resource growth rate per cell, and $r[c,t]$ is the level of resources in the cell at $t$.
The harvest of resources per cycle $H(t)$ is given as:
\begin{equation}
H(t) = \sum_{a=1}^{A(t-1)}r[c(a),t]+\delta_{p_{0}}(a)\sum_{a=1}^{B(t)}r[c(a),t]
\end{equation}
where $c(a)$ is the cell location of agent $a$ at the end of its action cycle, $A(t)$ are the agents alive at the end of cycle $t$, $B(t)$ are the agents born in cycle $t$, and $\delta_{p{0}}(a)$ is 1 if the duration of puberty is 0 and 0 otherwise.
The change of resources within the agent population $\Delta E_{p}(t)$ is written as
\begin{equation}
\Delta E_{P}(t) = H(t)-\sum_{a=1}^{A(t)}m(a) -\sum_{a=1}^{D(t)}[S_{a}(t)+m(a)]-\sum_{a=1}^{B(t)}[bc(a) -\delta_{p_{0}}(a)m(a)]
\end{equation}
where $m(a)$ and $bc(a)$ are the metabolism and birth cost of agent $a$ respectively, $D(t)$ are the agents who died in this cycle and $S_{a}$ is the surplus resources agent $a$ brought into this cycle. By requiring a conservation of resources, equation for the balance of resources  is given as:
\begin{equation}
	0 = \Delta E_{L}(t) + \Delta E_{P}(t) + \sum_{a=1}^{A(t)}[S_{a}(t)-S_{a}(t-1)]
\end{equation}

\appendix{ \textbf{Appendix B - Single Species Models from Mathematical Biology}}

A continuous homogeneous model of a single species population $N(t)$ was proposed by Verhulst in 1838 \cite{murray} :

\begin{equation}
	\frac{dN(t)}{dt}=rN(1-\frac{N}{K})
\end{equation}
where $K$ is the steady state carry capacity, $t$ is time, and $r$ is the intrinsic rate of growth. This model represents self-limiting, logistic growth of the population. This macroscopic model of a continuous, homogeneous population is quite descriptive and allows the exact solution
\begin{equation}
	N(t)=\frac{K}{[1+(\frac{K}{N_{0}}-1)e^{-rt}]}
\end{equation}
where $N_{0}$ is the initial population. 
While the continuous Verhulst Model fits the initial phase of growth well, it does not model oscillating population levels at the higher rates of intrinsic growth. 
A discrete form of the Verhulst process incorporating an explicit time delay $\tau$ in the self-limiting term was proposed by Hutchinson \cite{hutch} to account for delays seen in animal populations. The resulting discrete-delayed logistic equation \cite{wright}, often referred to as the Hutchinson-Wright equation \cite{kot} is then

\begin{equation}
	N(t+1)=[1+r-\frac{N(t-\tau)}{K}]N(t)
\end{equation}
This model's intrinsic growth rate with $\tau = 5$ captures the steady state, oscillating, and chaotic populations trajectories seen in the ABM with similar intrinsic growth rates.

\end{document}